\documentclass[12pt,ams,a4paper,nofootinbib]{revtex4}
\usepackage{amsmath}
\usepackage{graphicx}
\usepackage[usenames,dvipsnames]{color}

\newcommand{\re}{\ref}
\newcommand{\be}{\begin{equation}}
\newcommand{\ee}{\end{equation}}

\newcommand{\la}{\label}
\newcommand{\ber}{\begin{eqnarray}}
\newcommand{\eer}{\end{eqnarray}}

\newcommand{\bs}[1]{\ensuremath{\boldsymbol{#1}}}

\begin{document}

\title{Program to calculate coefficients of transformations 
between three--particle hyperspherical harmonics}

\author{  Victor D. Efros$^{a,b}$\footnote{E-mail address: v.efros@mererand.com}
  }

\affiliation{
$^{a}$National Research Centre "Kurchatov Institute", 123182 Moscow, Russia\\
$^{b}$National Research Nuclear University MEPhI, 115409 Moscow, Russia 
}
 
 
\begin{abstract}
\noindent
{\bf Abstract.} A program to calculate the three--particle 
hyperspherical brackets is presented. Test results are listed and it is seen that
the program is well applicable up to very high values of the hypermomentum 
and orbital momenta.
The listed runs show that it is also very fast. Applications of the brackets to
calculating interaction matrix elements and constructing hyperspherical bases for identical particles are
described. Comparisons are done with the programs published previously.

\end{abstract}

\bigskip

\maketitle
\begin{small}
\end{small} 

{\bf PROGRAM SUMMARY}

\begin{small}
\noindent
{\em Program Title:} HHBRACKETS                                         \\
{\em Licensing provisions: GPLv3}                                   \\
{\em Programming language:} Fortran-90                                  \\
{\em Nature of problem:}\\
When solving three--body problems, expansions of hyperspherical harmonics over 
harmonics similar in form but pertaining to different  sets of Jacobi vectors are required. 
A universal and fast routine that provides the coefficients of such expansions, called
hyperspherical brackets or Raynal--Revai coefficients, 
is needed to researchers in the field. The expansions are used
both to calculate interaction matrix elements and construct states (anti)symmetric with respect 
to particle permutations.\\
{\em Solution method:}\\
At the hypermomentum that is minimum possible at given Jacobi orbital momenta, 
hyperspherical brackets are calculated using an explicit expression that  includes only few
 summations. 
To calculate the brackets at larger hypermomenta, a recursion relation is 
employed.  It perfectly works up to very high  
hypermomenta.  Attention is paid to avoiding difficulties with large quantum numbers.
\end{small}

\section{Introduction}
	The first version of the program, named RRCOE,  
was written by the author in 1987  and  first used 
 (though not published)
in Refs.~\cite{dan} at studying
 halo nuclei. It was used afterwards by 
the author and his collaborators, 
e.g., \cite{gr}, as well as by other researchers.
 In the present version,
considerable modifications and improvements are done.
In particular, the program is made applicable  up to very large 
values of both the hypermomentum and orbital momenta.
Hyperspherical harmonics (HH) required in 
realistic three--nucleon calculations involve such large values.
The program is also made faster. Besides, the features of \mbox{Fortran--77} 
which at present are  
considered to be obsolete,
 are eliminated.

Expansions over HH are an efficient tool to solve three--nucleon problems. At present these
problems attract considerable attention in relation to nuclear interactions 
derived from Effective Field Theory.
Three--cluster nuclear systems are also intensively studied via 
solving three--body problems in the HH approach. 

\section{Hyperspherical brackets}

The three--particle Jacobi vectors
\be {\bs \xi}_1=\sqrt{\mu_{12}}({\bf r}_2-{\bf r}_1),
\qquad {\bs \xi}_2=\sqrt{\mu_{12,3}}[{\bf r}_3-
(m_1{\bf r}_1+m_2{\bf r}_2)/(m_1+m_2)]\la{j}\ee
are used below. Here ${\bf r}_i$ and $m_i$ are the particle positions and their masses,
 \mbox{$\mu_{12}=m_1m_2/[(m_1+m_2)m]$, and
\mbox{$\mu_{12,3}=(m_1+m_2)m_3/[(m_1+m_2+m_3)m]$} where  $m$ is
some reference mass}.  
In non--relativistic quantum mechanics, motion of the center of mass is 
separated and three--particle dynamics
problems pertain to the space spanned by the $|{\bs \xi}_1{\bs \xi}_2\rangle$
states. 

The three--particle HH we employ are denoted as $Y_{KLM}^{l_1l_2}$. The
quantum numbers labeling them are
the orbital momenta  $l_1$ and $l_2$ with respect to the  
${\bs \xi}_1$ and ${\bs \xi}_2$ vectors, the total orbital
momentum~$L$, its projection~$M$, and
the hypermomentum~$K$. The orbital momenta allowed at 
a given $K$ value are determined by the conditions that
$ l_1+l_2\le K$, parity of $l_1+l_2$ is the same 
as that of~$K$, and $|l_1-l_2|\le L\le l_1+l_2 $. The dimension of
the space of HH having the same~$K$, $L$, and $M$ values
is
\be N(K,L,M)=(i_1+1)(i_2+1),\quad i_1=L-\epsilon,\quad i_2=(K-L-\epsilon)/2\la{numb}\ee 
where $\epsilon=0$ or 1 when $K-L$ is even or odd, respectively.

HH depend on 5 angles
parametrizing  the 6--dimensional hypersphere $\xi_1^2+\xi_2^2\equiv\rho^2={\rm const}$. These
angles are denoted as~$\{\Omega\}$.
One has $d{\bs \xi}_1d{\bs \xi}_2=\rho^5d\rho d\Omega$. 
The
angles $\{\Omega\}$ are determined by the Jacobi vectors
and it is convenient here to represent the dependence on~$\{\Omega\}$ 
as that on the Jacobi vectors themselves,
$Y_{KLM}^{l_1l_2}=Y_{KLM}^{l_1l_2}({\bs \xi}_1,{\bs \xi}_2)$. 
The HH $Y_{KLM}^{l_1l_2}$ with different quantum numbers are orthogonal 
at integrating with $d\Omega$ over the hypersphere. The HH are taken normalized,
\[\int d\Omega Y_{KLM}^{l_1l_2*}({\bs \xi}_1,{\bs \xi}_2)Y_{K'L'M'}^{l_1'l_2'}
({\bs \xi}_1,{\bs \xi}_2)=
\delta(KLMl_1l_2,K'L'M'l_1'l_2'). \]

Consider the expansion of the HH $Y_{KLM}^{l_1l_2}({\bs \xi}_1,{\bs \xi}_2)$ 
over the HH of the same form but depending on 
${\bs \xi}_1'$ and ${\bs \xi}_2'$ vectors related with  ${\bs \xi}_1$ and ${\bs \xi}_2$ via
a pseudo orthogonal transformation
\be {\bs \xi}_1={\bs \xi}_1'\cos\varphi+{\bs \xi}_2'\sin\varphi,\qquad
{\bs \xi}_2={\bs \xi}_1'\sin\varphi-{\bs \xi}_2'\cos\varphi.\la{pse}\ee 
The HH $Y_{KLM}^{l_1l_2}({\bs \xi}_1,{\bs \xi}_2)$ when considered 
as functions of ${\bs \xi}_1'$ and ${\bs \xi}_2'$ are HH having the same
\mbox{$K$, $L$,} and $M$  values.
 Therefore, the expansion is of the form
\be Y_{KLM}^{l_1l_2}({\bs \xi}_1,{\bs \xi}_2)=
\sum_{l_1',l_2'}\langle l_1'l_2'|l_1 l_2\rangle^\varphi_{KL}
Y_{KLM}^{l_1'l_2'}({\bs \xi}_1',{\bs \xi}_2').\la{exp}\ee
The number of terms in the sum is given 
by Eq.~(\re{numb}).
 We address the coefficients 
\be \langle l_1'l_2'|l_1 l_2\rangle^\varphi_{KL}\la{rr}\ee
of this expansion. They are called hyperspherical brackets or 
Raynal--Revai coefficients, and they are real.

 The matrix (\re{rr}) is
symmetric, 
\be \langle l_1'l_2'|l_1 l_2\rangle^\varphi_{KL}=
\langle l_1l_2|l_1' l_2'\rangle^\varphi_{KL}.\la{sym}\ee
This follows from the fact that the matrix of the transformation  (\re{pse}) 
coincides with its inverse.
Indeed, then the matrix of the  expansion of $Y_{KLM}^{l_1l_2}({\bs \xi}_1',{\bs \xi}_2')$ 
over $Y_{KLM}^{l_1'l_2'}({\bs \xi}_1,{\bs \xi}_2)$
is also given by Eq.~(\re{rr}) when written in the same form.
But, at the same time, the latter matrix should be
 the transposed one to that of  Eq. (\re{rr}) since the transformation~(\re{pse}) conserves
the norms of~HH. 

The coefficients of the 
expansion of $Y_{KLM}^{l_1l_2}({\bs \xi}_1,{\bs \xi}_2)$ 
over $Y_{KLM}^{l_1'l_2'}({\bs \xi}_1',{\bs \xi}_2')$ in case of an orthogonal transformation
\be {\bs \xi}_1={\bs \xi}_1'\cos\varphi+{\bs \xi}_2'\sin\varphi,\qquad
{\bs \xi}_2=-{\bs \xi}_1'\sin\varphi+{\bs \xi}_2'\cos\varphi\la{ort}\ee
may obviously
 be written in terms of the above coefficients  (\re{rr}) as 
\[(-1)^{l_2}\langle l_1'l_2'|l_1 l_2\rangle^\varphi_{KL}.\]

\section{Applications of hyperspherical brackets}

Five angles parametrizing the 6--dimensional sphere may be chosen to be ones determining a 
  unit vector ${\bf n}_1$ in the direction of ${\bs \xi}_1$, a 
  unit vector ${\bf n}_2$ in the direction of ${\bs \xi}_2$, and  an 
extra angle $\theta$ such that $\xi_1=\rho\sin\theta$ and
$\xi_2=\rho\cos\theta$.  The HH $Y_{KLM}^{l_1l_2}({\bs \xi}_1,{\bs \xi}_2)$ 
are of the structure
\be [Y_{l_1}({\bf n}_1)Y_{l_2}({\bf n}_2)]_{LM}f_{Kl_1l_2}(\theta)\la{form}\ee
where $Y_{lm}({\bf n})$ are spherical harmonics and the 
brackets $[\ldots]$ represent the coupling to a total momentum. 
 The function $f_{Kl_1l_2}$ is as follows,
\[ f_{Kl_1l_2}(\theta)={\cal N}_{Kl_1l_2}\sin^{l_1}\theta\cos^{l_2}\theta
P^{(l_1+1/2,l_2+1/2)}_{(K-l_1-l_2)/2}(\cos2\theta)\]
where $P_n^{(\alpha,\beta)}$ is the Jacobi polynomial and ${\cal N}_{Kl_1l_2}$
is  the normalization factor assumed 
to be positive. 
The integration element $d\Omega$ 
is  $d{\bf n}_1d{\bf n}_2\sin^2\theta\cos^2\theta d\theta$. 

The relative--motion
kinetic energy operator of a three--particle system 
written in terms of the above Jacobi vectors   is
\[T= -(\hbar^2/2m)(\Delta_{{\bs \xi}_1}+\Delta_{{\bs \xi}_2}).\]
The Hamiltonian $H$ is $T+V$ and we
suppose that the interaction operator $V$ is of the form $V=V(12)+V(13)+V(23)$ where $V(ij)$ are interactions between
pairs of particles. 

Suppose here that the particles are not all identical. In the approach we discuss, 
basis functions contain the  HH $Y_{KLM}^{l_1l_2}({\bs \xi}_1,{\bs \xi}_2)$  coupled 
with the spin functions to a total momentum. 
The expansion coefficients depend on $\rho$.
In turn, they may be expanded over a suitable basis.
Kinetic energy matrix elements between the resulting basis functions 
are simple. In accordance with Eq. (\re{form}) 
such basis functions are also suitable to calculate  the matrix elements of the $V(12)$ interaction. 
And in order to calculate the matrix elements of the $V(13)$ interaction, the HH $Y_{KLM}^{l_1'l_2'}({\bs \xi}_1',{\bs \xi}_2')$ 
which are of
the  same form  but depend on the other Jacobi vectors,
\be {\bs \xi}_1'=\sqrt{\mu_{13}}({\bf r}_3-{\bf r}_1),\qquad {\bs \xi}_2'=\sqrt{\mu_{13,2}}[{\bf r}_2-
(m_1{\bf r}_1+m_3{\bf r}_3)/M],\la{j23}\ee
would be suitable.
The definition (\re{j23}) corresponds to the substitution \mbox{${\bf r}_2\leftrightarrow{\bf r}_3$, $m_2\leftrightarrow m_3$}
in the definition (\re{j}). The vectors (\re{j}) are expressed in terms of the vectors (\re{j23}) via a 
transformation of the form (\re{pse}).
  The transformation  matrix   is the
following,
\be \cos\varphi=\left[\frac{m_2m_3}{(m_1+m_2)(m_1+m_3)}\right]^{1/2},
\qquad
\sin\varphi=
\left[\frac{m_1(m_1+m_2+m_3)}{(m_1+m_2)(m_1+m_3)}\right]^{1/2}.\la{23}\ee
The calculation is then performed with the help of the corresponding expansion of Eq.~(\re{exp}) form. 

If any two of the three particles are not identical then the calculation of matrix elements of 
the $V(23)$ interaction is required in addition. The Jacobi vectors ${\bs \xi}_1'$ and ${\bs \xi}_2'$
suitable for this purpose are obtained from the
vectors (\re{j}) via the substitution 
${\bf r}_1\leftrightarrow{\bf r}_3$, $m_1\leftrightarrow m_3$.
The vectors (\re{j}) are expressed in terms of such vectors via a 
transformation of the form (\re{pse}) with the following transformation matrix,
\be \cos\varphi=\left[\frac{m_1m_3}{(m_1+m_2)(m_2+m_3)}\right]^{1/2},\qquad
\sin\varphi=-\left[\frac{m_2(m_1+m_2+m_3)}{(m_1+m_2)(m_2+m_3)}\right]^{1/2}.\la{13}\ee
The calculation
is then performed in the same way.

Hyperspherical brackets can also be applied to construct   an HH basis for identical particles.
Let us speak of the three--nucleon case. 
Basis functions include spin and isospin variables. These functions should be
antisymmetric with respect to particle permutations. 
They may be  constructed from HH and spin--isospin functions 
  both of which also transform  in a simple way
under permutations. In dynamics calculations, such basis functions provide maximum separation 
of space and spin--isospin degrees of freedom.

In the three--particle case, there exist three types of irreducible representations of the
corresponding permutation group. The representations of the first type are realized by states  
denoted as $\phi^s$ that are symmetric 
with respect to any permutations of particles. The representations   of the second type are realized by
states denoted as $\phi^a$ that are antisymmetric, i.e.
change their signs under particle
transpositions. Those of the third type are realized  by the so called states of mixed symmetry 
denoted as $(\phi'',\phi')$ that form a two--dimensional space 
invariant with 
respect to permutations. The states of mixed symmetry may be specified by the transformation formulae 
(see, e.g., \cite{ham}, Chapt. 7)
\be
(\hat{12})\left(\begin{array}{c}\phi'\\
\phi''\end{array}\right)=\left(\begin{array}{cc}-1&0\\0&1
\end{array}\right)\left(\begin{array}{c}\phi'\\
\phi''\end{array}\right),\qquad
(\hat{13})\left(\begin{array}{c}\phi'\\
\phi''\end{array}\right)=
\left(\begin{array}{cc}1/2 &
 -\sqrt{3}/2\\
-\sqrt{3}/2&-1/2
\end{array}\right)\left(\begin{array}{c}\phi'\\
\phi''\end{array}\right)
\la{ptr}\ee   
where $(\hat{{ ij}})$ are transpositions of particles $i$ and $j$.
The transformation matrices for other permutations may be obtained from here.

Correspondingly, basis HH having given symmetries and given $K$, $L$, and $M$ values are 
denoted as $Y^s_{KLM},(Y''_{KLM},Y'_{KLM})$, and $Y^a_{KLM}$. Basis spin--isospin functions of given symmetries
are denoted as
$\theta^s,(\theta'',\theta')$, and $\theta^a$.
The resulting basis antisymmetric states belong to one  of  
the following three types,
\be
Y^s_{KLM}\theta^a,\qquad 2^{-1/2}(Y'_{KLM}\theta''-Y''_{KLM}\theta'),\qquad Y^a_{KLM}\theta^s.\la{t3}
\ee 
(The hyperradial variable $\rho$ is invariant with respect to particle permutations so that 
in each case the $\rho$ dependence leads merely to  a
factor.)
 
Basis HH entering Eq. (\re{t3})
may be constructed via the application of the three--particle symmetrization (or Young)
 operators to the HH $Y_{KLM}^{l_1l_2}({\bs \xi}_1,{\bs \xi}_2)$.  These operators
may be represented in the following form,
\be
P^{[s]}=3^{-1}\left[1+(\hat{13})+(\hat{23})\right]P_+,\qquad
P^{[a]}=3^{-1}\left[1-(\hat{13})-(\hat{23})\right]P_-,\la{sm}
\ee
\be
P^{[m]}_{''('')}=3^{-1}\left[2-(\hat{13})-(\hat{23})\right]P_+,\qquad
P^{[m]}_{'('')}=3^{-1/2}\left[(\hat{23})-(\hat{13})\right]P_+,
\la{m1}
\ee
\be
P^{[m]}_{''(')}=3^{-1/2}\left[(\hat{23})-(\hat{13})\right]P_-,\qquad
P^{[m]}_{'(')}=3^{-1}\left[2+(\hat{13})+(\hat{23})\right]P_-
\la{m2}
\ee 
where
\be
P_{\pm}=2^{-1}[1\pm(\hat{12})].\la{ppm}
\ee
The transpositions $(\hat {ij})$ act as follows, 
$(\hat {ij})F({\bs \xi}_1,{\bs \xi}_2)=F((\hat {ij}){\bs \xi}_1,(\hat {ij}){\bs \xi}_2)$.
When  the operators (\ref{ppm}) are applied to the HH $Y_{KLM}^{l_1l_2}({\bs \xi}_1,{\bs \xi}_2)$ the result equals
one or zero depending on the parity of the angular momentum $l_1$. 
  Either pair of the operators $P^{[m]}_{''('')}, P^{[m]}_{'('')}$ or $P^{[m]}_{''(')}, P^{[m]}_{'(')}$ may be used
to construct mixed symmetry states $(\phi'',\phi')$.

Let us denote the operators (\re{sm})--(\re{m2}) as $P^{[f]}_{\mu(\mu')}$. 
 When these operators are applied to a complete set of  HH like
$Y_{KLM}^{l_1l_2}({\bs \xi}_1,{\bs \xi}_2)$  the sets of HH of given permutational symmetries arise which
are over-complete. 
It is possible to get bases in the spaces of the latter sets via orthogonalization of the HH thus obtained. 
Besides, matrix elements of kinetic energy between orthonormalized HH are simple. 
Therefore, one aims to construct orthonormalized
complete sets in the spaces of the $P^{[f]}_{\mu(\mu')}Y_{KLM}^{l_1l_2}({\bs \xi}_1,{\bs \xi}_2)$ states. 

Basis states forming
these sets are to be obtained
in the form of expansions over the initial  HH $Y_{KLM}^{l_1'l_2'}({\bs \xi}_1,{\bs \xi}_2)$. This is 
needed to calculate the interaction matrix elements and related to the fact that the
matrix elements of an interaction $V(12)+V(13)+V(23)$ between the states of Eq. (\re{t3}) type are the same as matrix elements
of $3\cdot V(12)$. 
A three--particle interaction may also consist of three terms that differ in numbering of particles and  are such 
that for calculating matrix elements of one of them, which  is sufficient, the mentioned expansion is suitable.

Thus, first one has to obtain expansions of the form
\be P^{[f]}_{\mu(\mu')}Y_{KLM}^{l_1l_2}({\bs \xi}_1,{\bs \xi}_2)=\sum_{l_1',l_2'}c_{KL}(l_1',l_2')
Y_{KLM}^{l_1'l_2'}({\bs \xi}_1,{\bs \xi}_2).\ee
According to Eqs.~(\re{sm})--(\re{m2}) the expansion coefficients $c_{KL}(l_1',l_2')$ are sums of the delta-- symbol contribution  
 and the contributions  
\be \int d\Omega Y_{KLM}^{l_1'l_2'*}({\bs \xi}_1,{\bs \xi}_2)Y_{KLM}^{l_1l_2}((\hat{ ij}){\bs \xi}_1,
(\hat{ ij}){\bs \xi}_2)
\la{mee}\ee
where $(\hat {ij})$ are transpositions $(\hat{13})$ and $(\hat{23})$. The $(\hat{23})$ contribution is
readily expressed in terms of  the $(\hat{13})$ contribution.

Calculating the latter, one may write
\be (\hat{13}){\bs \xi}_1={\bs \xi}_1\cos\varphi+{\bs \xi}_2\sin\varphi,\qquad 
(\hat{13}){\bs \xi}_2={\bs \xi}_1\sin\varphi-{\bs \xi}_2\cos\varphi.\la{trans}\ee
Therefore, in accordance with the definition of Eq. (\re{exp}) the matrix element  (\re{mee}) is equal to the hyperspherical
bracket
$\langle l_1'l_2'|l_1l_2\rangle_{KL}^{\varphi}$.

The transformation (\re{trans}) is reverse to the transformation of Eqs. (\re{pse}) 
and (\re{13}) with \mbox{$m_1=m_2=m_3$}.
The matrices of the transformation (\re{pse}) and its reverse are the same.
Hence
the matrix of the transformation (\re{trans})
is the matrix in Eq.~(\re{ptr}). 

In conclusion of this section 
let us comment on the orthogonalization procedure. It should be sufficiently stable if, as usual, the number (\re{numb}) of
HH with given $K$, $L$, and $M$ is not very high. Anyway, it may be performed with
the quadrupole precision. In the $(Y_{KLM}'',Y_{KLM}')$ case it is not required
to orthogonalize independently 
$Y_{KLM}''$ and $Y_{KLM}'$ states. 
Selecting linear independent states one needs to
 check whether their net number obtained coincides with that given
 by~Eq.~(\re{numb}). It is simpler to deal with the multiplicities 
of  the spaces of given permutational
symmetries  themselves when they known in advance. In the general case these multiplicities can be obtained as the traces of  
projection operators~\cite{fe}. And in the three--particle case we have obtained them analytically. We shall
list the formulae without derivation.

   To this aim, define three subsidiary functions $n_s(i)$, $n_m(i)$, and $n_a(i)$,  
\ber
n_s(i)=[(i+6)/6],\quad n_m(i)=i/3,\quad n_a(i)=[(i+3)/6]\qquad{\rm at}\,\,\, i({\rm mod}\,\, 3)=0, \nonumber\\
n_s(i)=[(i+2)/6],\quad n_m(i)=(i+2)/3,\quad n_a(i)=[(i-1)/6]\qquad{\rm at}\,\,\, i({\rm mod}\,\, 3)=1,\nonumber\\
n_s(i)=[(i+4)/6],\quad n_m(i)=(i+1)/3,\quad n_a(i)=[(i+1)/6]\qquad{\rm at}\,\,\, i({\rm mod}\,\, 3)=2.\la{sma}
\eer
Here $[\ldots]$ is the integer part of a number. (In all the cases
$n_s+2n_m+n_a=i+1$.)

Let $N_m(K,L,M)$, $N_s(K,L,M)$,   and $N_a(K,L,M)$ be, respectively,  the number 
of   independent subspaces $(Y_{KLM}'',Y_{KLM}')$ of mixed symmetry HH contained in the space of HH with 
given $K,L,M$,  the number  of
linearly independent HH with given $K,L,M$ that are 
symmetric with respect to particle permutations, and the number of such HH that are antisymmetric with respect to them.
One has $N_s+2N_m+N_a=N$
where $N(K,L,M)$ is given by Eq. (\re{numb}). The quantities 
$N_s$,  $N_m$, and $N_a$ sought for
are the following,
\ber
N_s(K,L,M)=n_s(i_1)n_s(i_2)+n_m(i_1)n_m(i_2)+n_a(i_1)n_a(i_2),\nonumber\\
N_m(K,L,M)=\nonumber\\
n_s(i_1)n_m(i_2)+n_m(i_1)n_s(i_2)+n_m(i_1)n_m(i_2)+n_a(i_1)n_m(i_2)+n_m(i_1)n_a(i_2),\nonumber\\
N_a(K,L,M)=n_s(i_1)n_a(i_2)+n_m(i_1)n_m(i_2)+n_a(i_1)n_s(i_2)\la{pro}
\eer 
where $i_1$ and $i_2$ are defined according to Eq. (\ref{numb}). 

The schemes of  three--body HH expansions may somewhat differ from those outlined above 
but the applications of the HH brackets are anyway the same.

\section{Relations to calculate the brackets}

These relations,  Eqs. (\re{1}) and (\re{2}) below, which were
 obtained in Ref.~\cite{se} are the following. It is convenient to deal with the modified brackets
\be [l_1'l_2'|l_1l_2]_{KL}^\varphi\equiv\langle l_1'l_2'|l_1 l_2\rangle^\varphi_{KL}A_{Kl_1l_2}/A_{Kl_1'l_2'}\la{mod}\ee
where
\be A_{Kl_1l_2}=\left[\left(\frac{K-l_1-l_2}{2}\right)!\left(\frac{K+l_1+l_2}{2}+1\right)!(K-l_1+l_2+1)!!(K+l_1-l_2+1)!!
\right]^{1/2}.\la{AA}\ee
First, the brackets with $K=l_1+l_2$ are calculated using 
the formula\footnote{ Denote the product $[l_1][l_2][l_1'][l_2']$
in Eq. (\re{1}) times the numerator of 
the ratio of factorials and double factorials
from the second line in this equation
  as $A$
and the denominator of the mentioned ratio as $B$. In the formula in Ref.~\cite{se}
the quantity $A/B$ was listed  as $A]^{1/2}/B$ which is an obvious misprint.}
\ber [l_1'l_2'|l_1l_2]_{l_1+l_2,L}^\varphi=[l_1][l_2][l_1'][l_2']
\nonumber\\
\times 
\frac{(l_1+l_2+1)!(l_1+l_2+L+1)!(l_1+l_2-L)!}{n_-!(n_++1)!(2n_-+2l_1'+1)!!(2n_-+2l_2'+1)!!}\,
2^{-(l_1'+l_2')}\sin^{n_-}\varphi\cos^{n_+}\varphi 
\nonumber\\
\times 
\sum_{m=m_{min}}^{m_{max}}(-1)^{\nu_4}\left[\prod_{i=1}^4\sqrt{(2\nu_i)!}/(\nu_i)!\right]
\left(\begin{array}{ccc}l_1-n_-&l_2-n_-&L\\l_1'-m&m-l_2'&l_2'-l_1'
\end{array}\right)
\tan^m\varphi\la{1}\eer
where  the notation like \mbox{$[l]=\sqrt{2l+1}$} is used, 
\mbox{$n_-=(l_1+l_2-l_1'-l_2')/2$}, \mbox{$n_+=(l_1+l_2+l_1'+l_2')/2$}, 
 the expression in the round brackets is 
the $3j$--symbol, and
\begin{eqnarray}
\nu_1=(l_1-n_-+l_1'-m)/2,\qquad\nu_2=(l_1-n_--l_1'+m)/2,\nonumber\\
\nu_3=(l_2-n_--l_2'+m)/2,\qquad\nu_4=(l_2-n_-+l_2'-m)/2.\la{nu}\end{eqnarray}
The summation goes between the limits 
\[m_{min}=\left|(l_1-l_1')-(l_2-l_2')\right|/2,
\qquad m_{max}={\rm min}\{l_1+l_1'-n_-,\,l_2+l_2'-n_-\}\]
within which the $3j$ symbol is different from zero. The summation variable 
takes only the values of the same parity as these limits. This is related to the requirement that 
the quantities (\re{nu}) should be integer.

Proceeding from the brackets of Eq. (\re{1}) the general type brackets are calculated with the help of the 
$K\rightarrow K+2$ recursion relation
\ber 
[l_1'l_2'|l_1l_2]_{K+2,L}^\varphi=\cos 2\varphi[l_1'l_2'|l_1l_2]_{KL}^\varphi+\sin \varphi\cos \varphi\nonumber\\
\times\left\{[l_1'-1l_2'-1|l_1l_2]_{KL}^\varphi\,\alpha_L(l_1',l_2')
-[l_1'+1l_2'+1|l_1l_2]_{KL}^\varphi\,\alpha_L(l_1'+1,l_2'+1)\right.\nonumber\\
\left.+[l_1'+1l_2'-1|l_1l_2]_{KL}^\varphi\,\beta_L(l_1',l_2')-[l_1'-1l_2'+1|l_1l_2]_{KL}^\varphi\,\beta_L(l_2',l_1')\right\}\la{2}
\eer
where
\ber\alpha_L(p,q)=\left[\frac{(p+q-L-1)(p+q-L)(p+q+L)(p+q+L+1)}{(4p^2-1)(4q^2-1)}\right]^{1/2},\nonumber\\
\beta_L(p,q)=\left[\frac{(q-p+L-1)(q-p+L)(p-q+L+1)(p-q+L+2)}{(2p+1)(2p+3)(4q^2-1)}\right]^{1/2}.\la{ab}\eer

Let us also mention that the oscillator  (or Moshinsky)
brackets can be calculated in a similar way \cite{efr1}.

\section{The program}

The double precision is set in the program.
Factorials and double factorials entering the above listed formulae are to be calculated as real numbers. 
The factors entering the product $[\ldots]$
in the third line in Eq. (\re{1}) are extracted from the array \mbox{$\sqrt{(2\nu)!}/\nu!$} calculated in advance.

 At calculating the second line factor in Eq. (\re{1}) the values of the products of factorials and double factorials
 entering it and even
those quantities themselves may become  larger than the maximum real number
allowed in a double precision calculation. To avoid this, the mentioned factor is calculated as $\exp(\ln A)$,
$A$ being this factor up to a sign. When a calculation is performed in this way, the large numbers 
cancel each other to a sufficient degree and the outcome is never too large or too small.
The quantity $\ln A$ includes the logarithms of factorials and 
double factorials. The arrays of these logarithms are calculated in advance. The ratio of the quantities (\re{AA})
is calculated similarly.

The point as to such large numbers arises also at
 calculating the $3j$ symbols entering Eq.~(\re{1}). We wrote a routine to calculate  $3j$ symbols utilizing the expression for them
 of 
the structure \mbox{$F^{1/2}\sum_n (-1)^nu_n^{-1}$} where $F$ is the ratio of products of factorials and
$u_n$ are also products of them, see, e.g., \cite{var}. 
The calculation is performed in the form \mbox{$\sum_n (-1)^n\exp[(1/2)\ln F-\ln u_n]$} which
 also in this case leads to cancellation to a sufficient degree of large numbers entering~$F$ and~$u_n$. 

We  tried also to proceed in another way constructing 
$F^{1/2}u_n^{-1}$ from the quantities like~$(i!)^{1/j}$ where $j$ is a sufficiently large integer. 
This makes possible to avoid the calculation of many exponentials. However, this made the net program faster
by less than only 30\%. Therefore, in order not to complicate the program we decided not to release this version.
We got rid of the difficulty also  via calculating the $3j$ symbols  in their primary form with the quadrupole
precision. However, in the case of
 brackets with large quantum numbers this increased the net computation time by 5 - 10 times and we abandoned also this
version.

The brackets are calculated with the help of the subroutine named\\
HHBRACKETS(K,L,L1,L2,CO,SI,DLFAC,DL2FAC,RFAC,N0,BRAC)\\
that is called from a main program.
HHBRACKETS returns the array BRAC of the brackets of Eq. (\re{rr}) with all $l_1'$ and $l_2'$ values allowed at given $K$, $L$,
$l_1$, and $l_2$. All the parameters of the subroutine but BRAC are input ones. CO and SI are $\cos\varphi$ and $\sin\varphi$. 
\mbox{DLFAC, DL2FAC}, and RFAC are 
arrays ranging from zero to $N_0$ 
 where $N_0$ is an arbitrary integer
larger than~$2K$.
The first two arrays contain, respectively,  the 
above mentioned quantities $\ln(m!)$ and  $\ln[(2m+1)!!]$, and the 
last one is the array of the above 
mentioned quantities $\sqrt{(2\nu)!}/\nu!$. To get these three arrays, a main program may
call for the appended
 small subroutine FACT(DLFAC,DL2FAC,RFAC,N0).

The subroutine HHBRACKETS we discuss calls for 
the function\\ WIGN(JJ1,JJ2,JJ3,MM1,MM2,DLFAC,N0)\\ that calculates the $3j$ symbols entering Eq. (\re{1}). It calls also for
the small routines AL and BE 	providing the quantities (\re{ab}). Thus the set of our routines consists of 
HHBRACKETS, WIGN, AL, BE, and FACT.

The output array of brackets 
reads as BRAC(M,N)  where $M$ and $N$ variables are related to the 
$l_1'$ and~$l_2'$ orbital momenta entering the brackets (\re{rr})
as follows,
\be l_1' = \frac{K-(L-\epsilon)}{2}+N-M,\qquad  l_2'=  \frac{K+L-\epsilon}{2}-M-N+2
\la{expr}\ee
where $\epsilon$ is as in Eq. (\re{numb}). The ranges of $M$ and~$N$  are the following,
\be 1\le M\le i_2+1,\qquad 1\le N\le i_1+1\la{mn}\ee 
where $i_1(K,L)$, and $i_2(K,L)$ are 
as in Eq. (\re{numb}).
It can be seen that the expressions (\re{expr}) set up a one--to--one correspondence between 
the $M$,~$N$ values of Eq. (\re{mn}) and all the 
allowed $l_1'$,~$l_2'$ values. The allowed
  $M$ and $N$   values are
independent of each other in the difference to the allowed $l_1'$ and $l_2'$ values.
When performing the summations at 
the calculations of matrix elements one may consider the brackets as depending on 
$M$ and~$N$ and employ the relations (\re{expr}).
One may  also use 
the expressions $M(l_1',l_2')$ and $N(l_1',l_2')$ reverse to those of Eqs. (\re{expr})
to extract brackets from the BRAC(M,N) array.

\section{tests}

Tests were conducted at $\cos\varphi=1/2$, $\sin\varphi=-\sqrt{3}/2$, see Eq. (\re{trans}).
The orthogonality relations 
\be \sum_{all\,\, l_1',l_2'}\langle l_1'l_2'|l_1 l_2\rangle^\varphi_{KL}\langle l_1'l_2'|l_3 l_4\rangle^\varphi_{KL}=
\delta_{l_1l_3}\delta_{l_2l_4}\la{te1}\ee
were tested for  various $K$ and $L$ values
of both parities at  $K$ up to 201 and $L$  up to 31. This presumably  covers all the possible applications.
The  relations with $l_3=l_1$ and $l_2=l_4$ were tested for all 
allowed $l_1$ and $l_2$ values. In each $K\simeq 200$, $L\simeq 30$ case, for example, the number of these relations 
is about 2700  and they involve about
7$\cdot10^6$ brackets thus controlled. In both $K=200$, $L=30$ and $K=200$, $L=0$ cases the largest among such relations
deviation from unity  was $3\cdot10^{-9}$. At $K=30$, $L=5$, for example,  the largest such deviation was $9\cdot10^{-14}$. 
 The accuracy is 
mostly determined 
by the accuracy at calculating the $3j$ symbols. 

At $K=200$ and
$L=4$ the zero values of the sum of Eq.~(\re{te1}) were 
reproduced as $10^{-14}-10^{-15}$ for some sets of randomly chosen allowed  $l_1$, $l_2$, $l_3$, and $l_4$ values
such that $l_3\ne l_1$ or $l_4\ne l_2$.
The symmetry of the calculated brackets was also verified at the same conditions. The two brackets coincided 
in 13 digits.

One more test was the following.
In the
$L=0$ case a complete set of HH that transform in a simple way under the coordinate transformation~(\re{pse}) or 
(\re{ort}) was found in a closed form~\cite{ga}. In Ref~\cite{efr}
it was (unexpectedly) found  that the transformation coefficients between that set and the $Y_{K,L=0}^{ll}$ set 
have the form of  the usual  Clebsch--Gordan coefficients
 of the $SO(3)$ group. 
In Ref.~\cite{se} this was employed to calculate the $L=0$ brackets also in another way using the HH of Ref.~\cite{ga}
as intermediate ones.
The corresponding expression includes a summation. At zero Jacobi orbital momenta, performing it directly one comes 
to the relation
\ber \langle 00|00\rangle^\varphi_{K,L=0}=\,\,\frac{2}{K+2}\frac{\sin(K+2)\varphi}{\sin 2\varphi}\la{simp}\eer
obtained in an implicit form in Ref.~\cite{fabr}.
The brackets entering this relation were tested at $K\simeq 200$ and they coincided with its right--hand side in 
14 digits.

Some results of the tests are presented in the appended file named "output". Tests at any quantum numbers
can be readily   performed with 
the appended program \mbox{TESTHHBRACKETS}. 

\section{Running Times}

The calculations below have been performed with a notebook Intel core~i5,~2.67~GHz~(2010). Brackets (\re{rr})
with {\it all}  $l_1$, $l_2$, $l_1'$, and $l_2'$ values allowed at given $K$ and $L$  were computed.
The numbers of these brackets 
are the squares of the numbers of states given by Eq. (\re {numb}).
 In the Table the  times are listed  to compute all these brackets 
(fourth column) along with the 
average times  per computing one of them (last column). The latter ones
are the former ones divided by the number of states listed in the the third column
(In all the cases except for the first one 
the same calculation was done repeatedly in a loop
 to have a detectable net CPU time value.)
\begin{table}[h]
\caption{Computing times in seconds}
\vspace{0.2cm}
\begin{tabular}{|c|c|c|c|c|}
\hline 
$K$& $L$& net number of brackets& net running time & average time per one bracket\\
\hline
\,200\,& \,30\,&7107556&48.7& $\,6.9\cdot10^{-6}\,$\\
\hline
200&0&10201&$2.62\cdot10^{-2}$&$2.6\cdot10^{-6}$\\
\hline
20&10&4356&$1.71\cdot10^{-3}$&$3.9\cdot10^{-7}$\\
\hline
20&0&121&$4.14\cdot10^{-5}$&$3.4\cdot10^{-7}$\\
\hline
6&2&81&$1.47\cdot10^{-5}$&$1.8\cdot10^{-7}$\\
\hline
6&1&9&$2.26\cdot10^{-6}$&$2.5\cdot10^{-7}$\\
\hline
6&0&16&$3.27\cdot10^{-6}$&$2.0\cdot10^{-7}$\\
\hline
\end{tabular}

\end{table}
In Refs.~\cite{kit1} and \cite{kit2} programs to calculate the HH brackets based on different algorithms were created. 
The program of Ref.~\cite{kit2} was tested there in the $K=6$ case.  60 brackets pertaining to $L=0$, 1, and 2
were calculated. The average computation time per one bracket equaled~2.4~s. In 
Ref.~\cite{kit1} where the brackets with 
$K=6$, $L=2$ were calculated the computation time was about the same. 
Our computation times for these cases are presented in
 the last three lines 
 in the Table.  
Basing on the flops, the increase in the speed of computation 
of the present  notebook with respect to
the VAX machine employed in Ref.~\cite{kit2} is about $3\cdot10^4$.
Very probably that it is about the same also with respect  
to the computer of  Ref.~\cite{kit1}. With this factor taken into account, one 
concludes that 
at $K=6$ the present
program is  more than
400 times faster as compared with the program of Ref.~\cite{kit2} and, very probably,
with that of Ref.~\cite{kit1}.
It is also not known
whether the codes~\cite{kit1,kit2} can produce sufficiently accurate results at
higher $K$ values.


\section{Acknowledgment}

Partial support from RFBR Grant No. 18--02--00778 is acknowledged.

\end{document}